\documentclass[square]{ws-procs11x85}
\usepackage{multicol,float} 

\def\apj{ApJ}
\def\jgr{J.~Geophys.~Res.}

\begin{document}

\title{3-d resistive MHD simulations of magnetic reconnection and the tearing mode instability in current sheets}

\author{G. C. Murphy$^*$}

\address{Laboratoire d'Astrophysique de Grenoble, CNRS, Universit\'e Joseph Fourier, Grenoble, France\\
$^*$E-mail: Gareth.Murphy@obs.ujf-grenoble.fr}

\author{R. Ouyed}

\address{ Department of Physics and Astronomy, University of Calgary,AB, Canada\\
E-mail: ouyed@phas.ucalgary.ca}

\author{G. Pelletier}

\address{Laboratoire d'Astrophysique de Grenoble, CNRS, Universit\'e Joseph Fourier, Grenoble, France\\
E-mail: Guy.Pelletier@obs.ujf-grenoble.fr}

\begin{abstract}
Magnetic reconnection plays a critical role in many astrophysical processes
where high energy emission is observed, e.g. particle
acceleration, relativistic accretion powered outflows, pulsar winds and probably
in dissipation of Poynting flux in GRBs. The magnetic field
acts as a reservoir of energy and can dissipate its energy to thermal and
kinetic energy via the tearing mode instability. We have performed 3d
nonlinear
MHD simulations of the tearing mode instability in a current sheet. Results from
a temporal stability analysis in both the linear
regime and weakly nonlinear
(Rutherford) regime are compared to the numerical simulations. We observe
magnetic island formation, island
merging and oscillation once the instability has saturated. The growth in the
linear regime is exponential in agreement with linear theory. In
the second, Rutherford regime the island width grows linearly with time. We find
that thermal energy produced in the current sheet strongly
dominates the kinetic energy. Finally preliminary analysis indicates a P(k) ~
4.8
power law for the power spectral density which suggests that
the tearing mode vortices play a role in setting up an energy cascade.
\end{abstract}


\bodymatter

\begin{multicols}{2}
\section{Introduction}
 Magnetic reconnection plays a critical role in many astrophysical processes, e.g. particle acceleration \cite{Phan}, accretion disks \cite{Hawley} and solar flares \cite{Shibata}. 
It is also important in laboratory fusion devices such as tokamaks.  
Magnetic reconnection is a topological change in the field which violates the frozen-flux condition of ideal magnetohydrodynamics (MHD). 
If a magnetic field can leak across the plasma it can reach a lower energy state - in the case of a current sheet it can undergo ``tearing'' into filaments or magnetic islands.  
A current layer of thickness $a$ may dissipate on timescales shorter than the resistive timescale $a^2/\eta$ due to the tearing mode instability (hereafter TMI).  
The TMI was first discovered in tokamaks and stellerators and has been extensively studied since the pioneering works \cite{Furth} and \cite{Rutherford} (see also \cite{White}, \cite{PriestForbes}, \cite{Biskamp}). In this paper we present numerical simulations of the dissipation of current sheets and the formation of magnetic islands due to the tearing mode instability.  In Section \ref{Analysis} we remind the reader of the predictions of the linear and weakly nonlinear analyses of Furth et al. \cite{Furth} and Rutherford \cite{Rutherford}.  In Section \ref{Method} we present our numerical method. In Section \ref{Results} we compare the results with linear and weakly nonlinear theory.  In Section \ref{Turbulence} we present power spectra derived from 3d simulations.  In Section \ref{Discussion} we discuss the pitfalls of simulations where reconnection is not properly tracked and the implications for reconnection-driven turbulence.

\section{Theory}
\label{Analysis}

\subsection{The linear regime}

In the linear analysis it is assumed that resistivity is only important within
the current layer.
Separate solutions may then be derived for the exterior, ideal MHD region and
the
interior, resistive region.
Using asymptotic matching \cite{Furth} derived stability range and
growth rates for the TMI in the
inviscid linear regime, neglecting the effects of compressibility.
The range of the instability is $\alpha < 1, \alpha=ak $, where k is the wave
number and $a$ is the current sheet width.

The growth rate is:
\begin{equation}
\gamma \sim \alpha^{-\frac{2}{5}} S^{ \frac{2}{5}} / \tau_R
\end{equation}

The linear growth rate is a function of the mode number, $k$, the current sheet
width, $a$, and the Lundquist or magnetic Reynolds number, $S$.
The equations are valid for $\alpha > S^{-\frac{1}{4}}$.
The system stays linear for as long as the magnetic island does not exceed the
width of the current sheet.

\subsection{Non-linear regime}
Rutherford \cite{Rutherford} studied the evolution of the TMI in the weakly
nonlinear
regime, by considering the effect of second order eddy currents on the current
sheet.
\cite{Rutherford} found in the nonlinear regime that the island width grows as t,
instability growth rate slows down from exponential to $t^2$, and that the
critical amplitude where the linear
solution ceases to be valid is:
\begin{equation}
B_{max}=\frac{\sqrt{2\eta\rho\gamma}}{\alpha}
\end{equation}

\begin{table}[H] 
\tbl{ Growth Rates and Saturation Times for Lundquist number 2400}
{\begin{tabular}{@{}lcccr@{}}\toprule
 $\alpha$ &\multicolumn{2}{l} {Linear Growth Rate [$\tau_A^{-1}$]
 }&\multicolumn{2}{l}{Saturation [$\tau_A$] }\\\colrule
 & Theory & Sim & Theory & Sim \\ 
 0.01 & 0.0590 &  0.0528 & 153.233 & 170 \\ 
 0.04 & 0.0339 &  0.0484 & 394.980 & 250 \\
 0.3 & 0.0151 &  0.01276 & 508.370 & 520 \\\botrule
\end{tabular}}\label{aba:tbl1}
\end{table}
\subsection{Previous work}

More recent work has concentrated on including the physics of the Hall effect
using multifluid or Hall MHD and electron inertia, using PIC techniques.
Significant increase in reconnection rates is found in these studies \cite{Birn}. However in these works the authors do not compare against analytical theory nor is the 3D power spectrum calculated.

\section{Numerical setup}
\label{Method}

We perform first 2D and then 3D simulations in resistive MHD using the
astrophysical code PLUTO.
%
%
%
%
%
%
%
%
%
%

As in any numerical code some numerical resistivity is present. In the linear
analysis numerical resistivity has been assumed to be small in comparison with
the physical resistivity.
Our initial mean field is a Harris current sheet of the form
$B_y=\mathrm{tanh}(x)$.
We initialise the field in the magnetic vector potential $A_z=\mathrm{ln}
|\mathrm{cosh(x})|$
and take the curl to derive $\mathbf{B}.$
To this we add a perturbation of the form cos(kx) in 2d and cos(kx)cos(ky) in
3d.
Free parameters are $\alpha$, $\beta=\frac{p}{B^2}$ and the resistivity $\eta$.
Zero-gradient Neumann-like ``outflow'' boundary conditions often produce
subsonic reflection for
compressible flows \cite{Roache,Colonius}. We use
a coarse mesh boundary which reduces this effect.

\section{Simulation Results}
\label{Results}
\subsection{2D}
\begin{figure}[H]
\centerline{\psfig{file=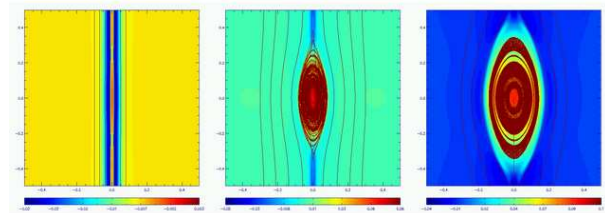,width=8cm}}
\caption{\label{fig:islforma0.01}
Formation of magnetic islands: magnetic field lines and density colourmap are
shown at times: 245, 365, 490.
 }
 \end{figure}
 \begin{figure}[H]
\centerline{\psfig{file=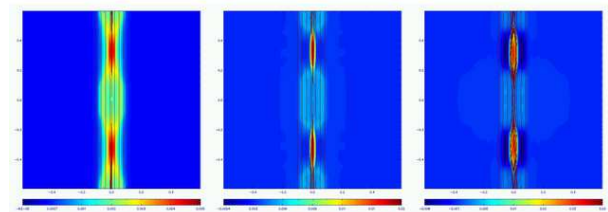,width=8cm}}
 \caption{\label{fig:islforma0.3}
 Same as Figure \ref{fig:islforma0.01} but for $\alpha=0.3$
  }
  \end{figure}
We performed 2d and 3d simulations of current sheets for different initial
perturbations $k=0.01,0.04,0.3,0.6$. We tracked the growth rate of the
cross-sheet magnetic field $B_x$.
In the simulations an initial transient period, $t_{transient}\sim 100 \tau_A$
roughly corresponding to half the crossing time of the domain ($t=112 \tau_A$)
was observed
before the linear mode was established.
As the tearing mode grows linearly, field lines reconnect, new separatrices are
formed and magnetic islands grow in size, eventually exceeding the size of the
original current sheet.
In Figures \ref{fig:islforma0.01} and \ref{fig:islforma0.3} the time evolution
of
magnetic islands for $\alpha=0.01,0.3$ are plotted against time in units of the Alfv\'en crossing time, $\tau_A$.

  \begin{figure}[H] 
\centerline{\psfig{file=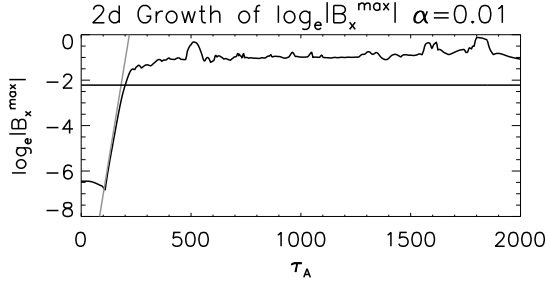,width=8cm}}
  \caption{\label{fig:2drecoa0.01}
  Growth of log of maximum $B_x$ plotted against time. The linear growth rate is
  plotted and the critical value for saturation.
   }
	\end{figure}
	 \begin{figure}[H]
\centerline{\psfig{file=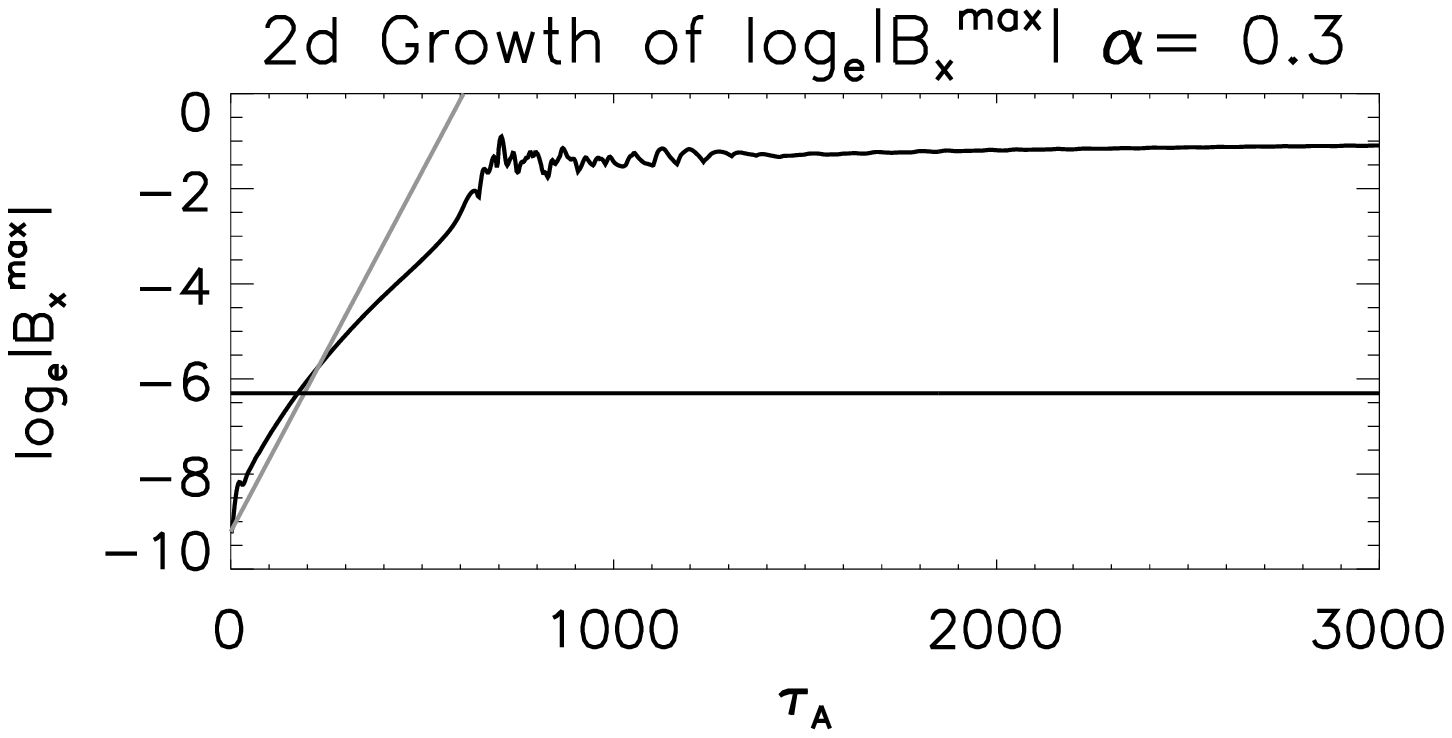,width=8cm}}
	 \caption{\label{fig:2drecoa0.3}
Same as Fig. \ref{fig:2drecoa0.01} but for $\alpha=0.3$.
	  }
	  \end{figure}

\begin{figure}[H]
\centerline{\psfig{file=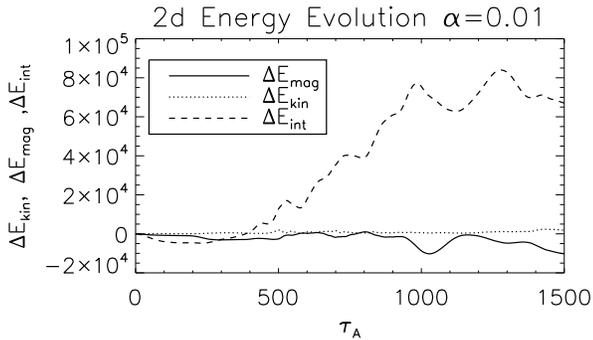,width=8cm}}
\caption{\label{fig:2denergya0.01}
Time evolution of change in kinetic energy, magnetic energy and internal energy.
Most of the magnetic energy is converted into thermal energy.
 }
\end{figure}



  \subsubsection{Agreement with linear analysis}
  In Figures \ref{fig:2drecoa0.01} and
  \ref{fig:2drecoa0.3} the growth rate in cross-sheet magnetic field, $B_x$ for
  different exciting modes of the TMI is plotted against time. In all cases a
  linear regime is found with slopes near the analytical predictions of Furth et
  al. \cite{Furth}.
 
\begin{figure}[H]
\centerline{\psfig{file=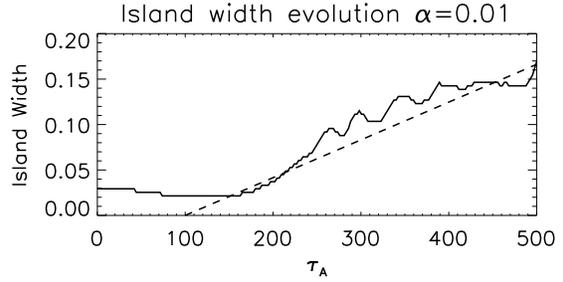,width=8cm}}
\caption{\label{fig:2disla0.01}
Magnetic island width evolution plotted against time in units of $\tau_A$ for
$\alpha=0.01$. The width is estimated using
the full-width half maximum of the thermal pressure.
It remains approximately constant in the linear phase and transfers to a linear
growth in $t$ once in the Rutherford regime.
 }
\end{figure}


\begin{figure}[H]
\centerline{\psfig{file=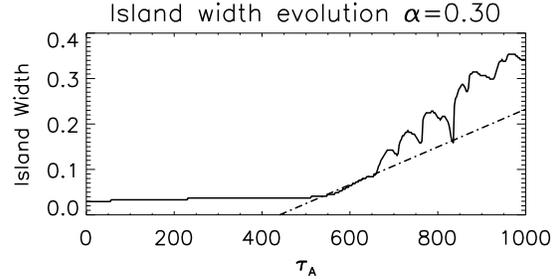,width=8cm}}
\caption{\label{fig:2disla0.3}
Same as \ref{fig:2disla0.01} but for $\alpha=0.3$.
 }
\end{figure}

\subsubsection{Agreement with nonlinear analysis}

We find plotting the island width against time that it remains small in the
linear regime and grows as $t^2$ in the Rutherford regime.
We can estimate the time of transition from linear to Rutherford from both the
$B_x$ plots and the island width plots, however the island width plots provide a
more accurate estimate.
The time at which the system becomes nonlinear is tabulated in Table
\ref{aba:tbl1}.
The magnetic island width can be derived from eqn 18. in \cite{Rutherford,White} and can be written in approximate form as
\begin{equation}
w_I \sim \frac{\tilde{t} - t_{d}}{S}, 
\end{equation}
where $t_{d}$ accounts for the total delay (transient and time spent in linear
regime) in our simulations before entering the Rutherford regime.

\begin{table}
\caption{Growth Rates and Saturation Times for Lundquist number 2400}
\begin{center}
 \begin{tabular}{ |l |l| l| l| l| }
 \hline
 $\alpha$ &\multicolumn{2}{l|} {Linear Growth Rate [$\tau_A^{-1}$]
 }&\multicolumn{2}{l|}{Saturation [$\tau_A$] }\\
 \hline
 & Theory & Sim & Theory & Sim \\
 0.01 & 0.0590 &  0.0528 & 153.233 & 170 \\
 0.04 & 0.0339 &  0.0484 & 394.980 & 250 \\
 0.3 & 0.0151 &  0.01276 & 508.370 & 520 \\
 \hline
 \end{tabular}
 \label{tab:2dtable}
\end{center}
\end{table}

\section{3D Power spectra}
\label{Turbulence}

\begin{figure}[H]
\centerline{\psfig{file=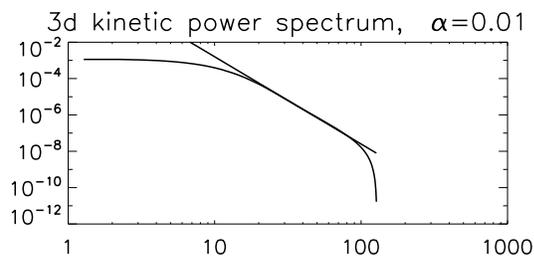,width=8cm}}
\caption{\label{fig:psdmagnetic}
3d kinetic power spectrum at time $t=295\tau_A$ with overplotted $k^{-4.8}$
 }
\end{figure}

In Figure \ref{fig:psdmagnetic} we plot the power spectrum for $\beta=2$ 3d current sheet. The slope of $P(k)$ is $-4.8$, within the range of values found by \cite{Vestuto} for $\beta=1$ and $\beta=\infty$. \cite{Vestuto} note that in their driven, supersonic turbulence simulation they have a constant beta and a constant mass-to-flux ratio ``modulo the numerical reconnection effect''. We obtain the same power spectrum scaling without any added velocity perturbations.

\section{Discussion and Conclusion}
\label{Discussion}

\subsection{Consequences for driven-turbulence simulations}

Our results show that even without either a driven turbulence mechanism, or an
initially turbulent velocity spectrum it is possible for the vortices generated
by the TMI to mimic the power spectrum seen in simulations of turbulence. This
can have serious consequences for MHD simulations where the resistivity is not
explicitly constrained (i.e. numerical resistivity plays a role), since there 
will be inevitably be some numerical-reconnection driven vortices present in the simulation. However using an explicit resistivity, the contribution from 
reconnection can be estimated using the formulae in \cite{Furth,Rutherford}.

Finally, we propose as a benchmark for resistive MHD codes the tearing mode instability test,
which can be compared with both analytical results for both the linear and
nonlinear regimes, as well as with laboratory experiments. The test may also
prove useful as a way of quantifying the effects of numerical resistivity in an MHD code.
 
\section*{Acknowledgements}
G.C.M. would like to acknowledge funding from the Agence Nationale de Recherche, the University of Calgary, and the Dublin Institute for Advanced Studies.

\bibliographystyle{ws-procs11x85}

\end{multicols}

\end{document}